\documentclass[10pt]{iopart}
\usepackage{graphicx}
\usepackage{citesort}

\begin{document}

\title{Schottky barrier and contact resistance of InSb nanowire field effect transistors}

\author{Dingxun Fan$^1$, N~Kang$^1$, Sepideh~Gorji~Ghalamestani$^2$, Kimberly~A~Dick$^2$ and H~Q~Xu$^{1,2}$}
\address{$^1$ Department of Electronics and Key Laboratory for the Physics and Chemistry of Nanodevices, Peking University, Beijing 100871, China}
\address{$^2$ Division of Solid State Physics, Lund University, Box 118, S-221 00 Lund, Sweden}
\eads{\mailto{nkang@pku.edu.cn}, \mailto{hqxu@pku.edu.cn}}

\vspace{10pt}
\begin{indented}
\item[]\today
\end{indented}

\begin{abstract}
Understanding of the electrical contact properties of semiconductor nanowire (NW) field effect transistors (FETs) plays a crucial role in employing semiconducting NWs as building blocks for future nanoelectronic devices and in the study of fundamental physics problems. Here, we report on a study of the contact properties of Ti/Au, a widely used contact metal combination, to individual InSb NWs via both two-probe and four-probe transport measurements. We show that a Schottky barrier of height $\Phi_{\rm{SB}}\sim20\ \rm{meV}$ is present at the metal-InSb NW interfaces and its effective height is gate tunable. The contact resistance ($R_{\rm{c}}$) in the InSb NWFETs is also analyzed by magnetotransport measurements at low temperatures.  It is found that $R_{\rm{c}}$ at on-state exhibits a pronounced magnetic field dependent feature, namely it is increased strongly with increasing magnetic field after an onset field $B_{\rm{c}}$. A qualitative picture that takes into account magnetic depopulation of subbands in the NWs  is provided to explain the observation.
Our results provide a solid experimental evidence for the presence of a Schottky barrier at Ti/Au-InSb NW interfaces and can be used as a basis for design and fabrication
of novel InSb NW based nanoelectronic devices and quantum devices.
\end{abstract}

\vspace{2pc}
\maketitle

\section{Introduction}

The operation of a field-effect transistor (FET) relies on a delicate control of the current flow through the conduction channel  by gate voltage. Integration of billions of tiny FETs on a chip constitutes  the cornerstone of modern electronics. As further scaling in the FET feature size becomes technically and principally challenging, researchers are seeking for alternatives with unique electronic properties to bypass the bottleneck. Of particular interest among them is to employ semiconducting nanowires (NWs) as conduction channels because of their inherent advantages in making junctionless and nonplanar wrap-gate devices~\cite{Colinge2010,Heike2014}. The presence of a Schottky barrier (SB) at a metal-semiconductor NW interface, however, hinders efficient charge carrier injection into the semiconductor NWs and limits the device performance. In conventional silicon technology, this problem is tackled by degenerately doping in the contact area~\cite{Sze2007}. Although the properties of electrical contacts to NWs can be significantly different from their bulk counterparts due to geometrical aspects and electrostatics~\cite{Francois2011}, extensive efforts have been made to optimize the carrier injection into NWs by introducing doping either  \emph{in situ} at  growth stage~\cite{Anil2012,Lu2007,Chen2012,Chen2013} or \emph{ex situ} via annealing during device fabrication~\cite{Weber2006,Wu2004}. An fundamentally different approach is, however, to utilize rather than circumvent the SBs for the realization of novel nanodevices and applications~\cite{Larson2006,Appenzeller2008,Massimo2012,Andre2012,Andre2013,Sebastian2014}. Along with this line are the experimental and theoretical research efforts to directly measure the metal-NW SB heights~\cite{Freitag2001,Martin2011} and to understand the carrier transport mechanism at the interfaces~\cite{Heinze2002,Appenzeller2002,Appenzeller2004}. It is anticipated that an additional contact resistance is present at a metal-NW interface due to the formation of subbands in the NW~\cite{Datta1995}. Usually, this part of the resistance is small compared with the resistance of the conduction channel at high temperatures. But it can become the dominant source of the resistance in a device with a ballistic NW channel at low temperatures~\cite{Wees1988,Wharam1989,Picciotto2001}. Therefore, a combined study of the Schottky barrier height $\Phi_{\rm{SB}}$ and the contact resistance $R_{\rm{c}}$ is desired in the evaluation of the properties of metal-NW interfaces.

Recently, InSb NWs have stimulated intense research interest. Due to the smallest band gap and the highest electron mobility in bulk InSb among all binary III-V semiconductors, InSb NWs own considerable potential for applications in low-power electronics and infrared optoelectronics~\cite{Borg2013,Nilsson2012}. Current experimental research focuses on employing of strong spin-orbit interaction strengths and large Land\'{e} g factors in InSb NWs~\cite{Nilsson2009,Perge2012,Fan2015} for detection and manipulation of Majorana fermions in solid state~\cite{Mourik2012,Deng2012,Churchill2013,Deng2014} and for implementation of spin-based quantum computation by means of electrical gates~\cite{Perge2012,vandenBerg2013}. Notably, most aforementioned devices comprise at least one normal metal-InSb NW contact. Before moving forward to more complex devices, it is necessary to understand solidly the transport properties at the interfaces between metal electrodes and semiconductor NWs. In the case of InAs NWs, it has been reported that the Fermi level is pinned to the conduction band at the contacts due to the formation of charge accumulation layers at NW surfaces~\cite{Noguchi1991,Olsson1996}. A recent study has, however, shown that SBs can be present at the contacts of the InAs NWFET with a small NW diameter due to quantum confinement~\cite{Razavieh2014}. It is thus natural to ask whether a SB exists at a metal-InSb NW interface, given the fact that the electron effective mass in InSb ($m_{\rm{e}}^{*}=0.015\ m_{\rm{e}}$) is smaller than InAs, and a quantitative study of the transport properties at metal-InSb NW interfaces is therefore required for the  realization of a rational control of the carrier injection into InSb NW channels. It should be noted that the impact of SBs on the performance of narrow band gap NWFETs has been discussed in a recent work~\cite{Zhao2012}. But a direct extraction of the SB height $\Phi_{\rm{SB}}$ still remains absent. In addition,  the contact resistance $R_{\rm{c}}$ has either not been included~\cite{Nilsson2010,Wang2011,Yao2012} or assumed to stay at a fixed value~\cite{Weperen2012,Gul2015} in previous studies of InSb NWFETs. The effect of $R_{\rm{c}}$ as a function of magnetic field even remains mostly unexplored in electrical transport measurements of InSb NWFETs.

Here, we present a detailed study of carrier transport and the effect of quantum confinement on the contact resistance in Ti/Au-contacted InSb NWFETs. A small SB of height $\Phi_{\rm{SB}}\sim20\ \rm{meV}$ is determined from the temperature-dependent transfer characteristics of the NWFETs. The effective barrier height is also extracted and is found to be gate tunable. This property enables us to further investigate the contact resistance at the on-state of the devices  arising from the formation of quasi-one-dimensional subbands in the NWs. The measured contact resistance shows a device-to-device variance of $3.63-6.81\ \rm{k\Omega}$ at on state. Applying a magnetic field markedly increases the contact resistance of InSb NWFETs at on state. The effect is interpreted  within a picture that captures the electron cyclotron radius as the characteristic length scale. The picture is further verified by transport measurements at successive magnetic fields.

\section{Materials and methods}

The InSb NWs employed in this work were grown on InAs (111)B substrates in a metal-organic vapor phase epitaxy reactor (Aixtron 200/4) operated at 100 mbar using hydrogen carrier gas with a total flow of 13\ L/min. Gold aerosol nanoparticles with selected diameters of 40 and 50~nm, and a total nanoparticle density of 1 $\mu m^{-2}$ were employed as seeds. Substrates were first annealed at 550~$^{\circ}$C for 7 minutes under a flow of arsine (AsH$_3$) with molar fraction $1.54\times10^{-3}$ to remove surface oxide. InAs stem nanowires were then grown for 7 minutes at 450~$^{\circ}$C using trimethylindium (TMIn) with a molar fraction of $4.81\times10^{-6}$ and AsH$_3$ molar fraction of $3.85\times10^{-4}$. Finally, InSb nanowire growth was initiated by simultaneously switching off the AsH$_3$ flow and switching on a flow of trimethylantimony (TMSb) with a molar fraction of $4.53\times10^{-5}$. InSb nanowires were grown for 30 minutes and had an average length of 2.5 $\mu m$, as determined by scanning electron microscopy inspection (Zeiss Leo 1560 operated at 10 kV). Depending on the sizes of catalytic gold particles, these InSb NWs have diameters in a range of $90-110\ \rm{nm}$ and are single zincblende crystals as determined from transmission electron microscopy.

As-grown InSb NWs were mechanically transferred onto a degenerately doped Si substrate, used as global back gate, with a 105-nm-thick capping layer of SiO$_2$. Selected NWs were located relative to predefined markers on the substrate. Standard electron beam lithography (EBL) process was carried out to define the contact areas on the selected NWs. A  Ti/Au (5/120 nm) metal film was then deposited in an electron-beam  evaporator on the exposed areas of the NWs. Here, a thin layer of Ti was used to promote metal adhesion to the NWs and the substrate. Prior to metal evaporation, a chemical etching in a H$_2$O-diluted (NH$_4$)$_2$S$_x$ solution at 40~$^{\circ}$C was used to remove the surface oxide layer on the NWs~\cite{Suyatin2007}. After lift-off in acetone, as-fabricated devices were immediately stored in a dark vacuum box to prevent the contacts from degradation. A description of the details about the devices presented in this work are listed in \textit{Supplementary data Table~S1}.

Figures 1a and 1b show false-color coded SEM images of a two-probe and a four-probe InSb NWFET.  In the measurements for the two-probe device (Figure 1a), a dc voltage $V_{\rm{ds}}$ is applied between the source (S) and the drain (D) electrode, and the current $I_{\rm{ds}}$ is recorded with the same electrodes.
In this case, it is inevitable that part of the applied voltage drops at the two contacts, with each characterized by a contact resistance $R_{\rm{c}}$.
The measured resistance is given by $R=V_{\rm{ds}}/I_{\rm{ds}}=2R_{\rm{c}}+R_{\rm{NW}}$, where the contact resistances at the two contacts are assumed to be the same and $R_{\rm{NW}}$ is the resistance of the NW channel. While in the measurements for the four-probe device (Figure 1b), a dc current $I_{\rm{ds}}$ is applied to the NW through the two outer electrodes 1 and 4, and the voltage drop $V$ between the two inner probes 2 and 3 is recorded.
In this way, the measured resistance, $R=V/I_{\rm{ds}}=R_{\rm{NW}}$, reflects mostly the NW channel resistance and thus the four-probe measurement method provides a direct measure of the channel transport properties.

\begin{figure*}
  \centering
  \includegraphics[width=5.6in]{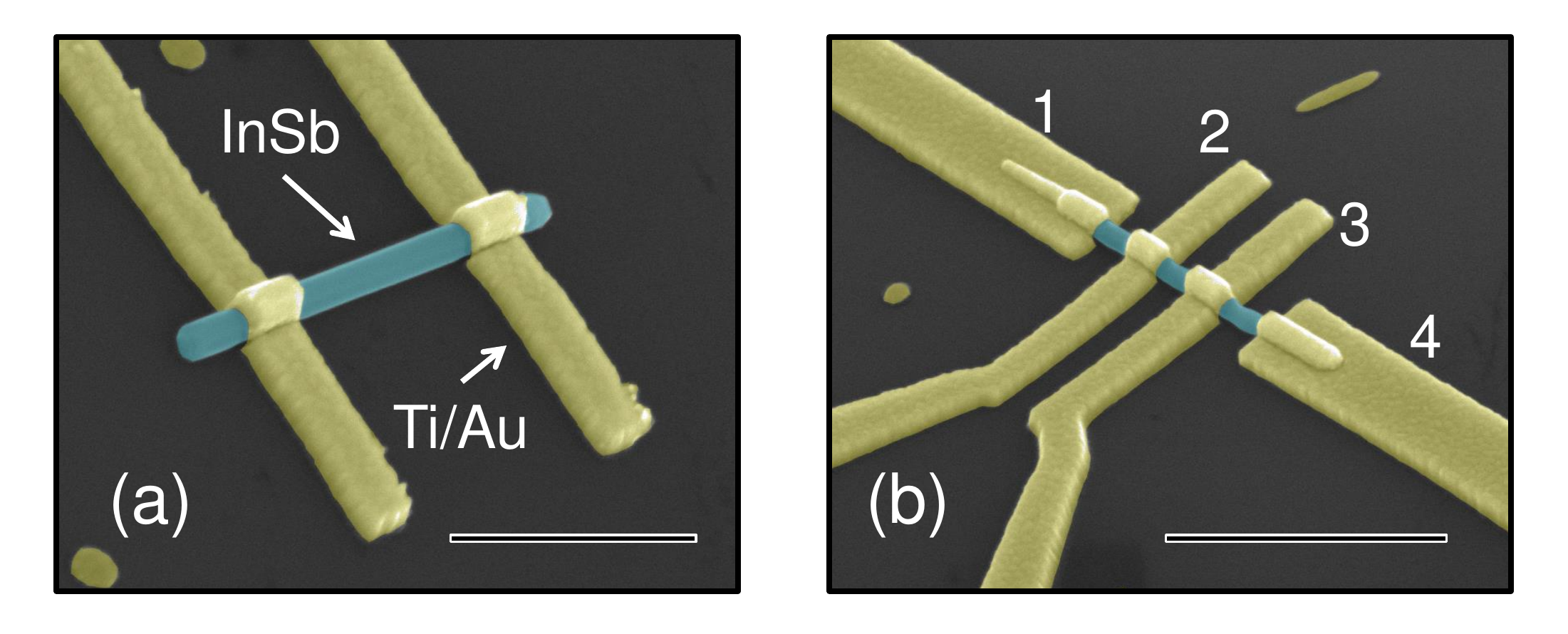}
  \caption{Representative false-colored SEM pictures showing a $40^{\circ}$ tilted view of (a) a typical two-probe and (b) a typical four-probe InSb NWFET fabricated for this work. The scale bars correspond to $1\ \rm{\mu m}$ in both pictures. The NWs employed in the device fabrication have a diameter in a range of $90-110\ \rm{nm}$. The two-probe and four-probe devices are used to extract the Schottky barrier height and the contact resistance at the Ti/Au-InSb NW interfaces, respectively.}
  \label{Figure 1}
\end{figure*}

\section{Results and discussion}

\begin{figure*}

  \centering
  \includegraphics[width=5.6in]{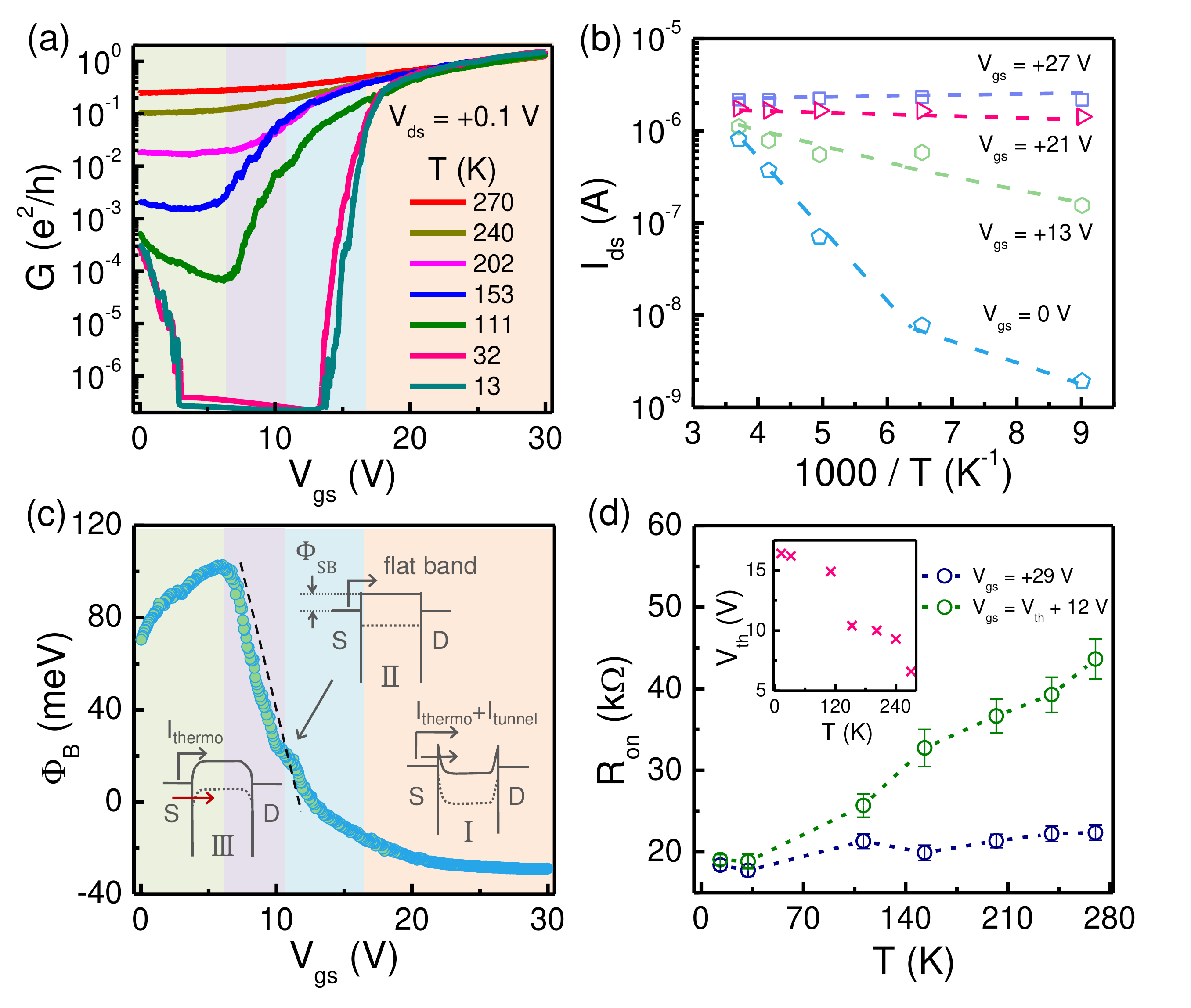}
  \caption
  {Schottky barrier height $\Phi_{\rm{SB}}$ at Ti/Au-InSb nanowire interfaces. (a) Transfer characteristics of a two-probe device (Device 1) measured at different temperatures at a source-drain bias voltage of $V_{\rm{ds}}=0.1\ \rm{V}$. (b) Arrhenius plot converted from the measurements shown in (a) at various gate voltages $V_{\rm{gs}}$. (c) Effective barrier height $\Phi_{\rm{B}}$ extracted from the Arrhenius plots in the high temperature region shown in (b). The insets I, II, and III show the energy band alignments for the device at on state, the flat-band case, and the case for the emergence of hole current (indicated by a red arrow). Note that the schematics in the insets are all drawn at the zero-bias condition for simplicity. In the measurements, a finite bias voltage of $V_{\rm{ds}}=0.1\ \rm{V}$ is applied to eliminate the effect of the SB on the drain side. (d) Resistance $R_{\rm{on}}$ of the device at on state measured as a function of temperature $T$ at a constant gate voltage of $V_{\rm{gs}}=29$ V (blue circle) and a constant gate voltage interval of 10 V relative to the threshold voltage $V_{\rm{th}}$ (green circle). Each data point is obtained by averaging over a gate voltage range of $5\ \rm{V}$ with its standard error indicated by a vertical bar. The inset shows $V_{\rm{th}}$ as a function of temperature $T$.}
  \label{Figure 2}
\end{figure*}

Figure~2 shows the transfer characteristics of the device shown in Figure~1a (Device 1) measured in the two-probe configuration at different temperatures and the extracted SB height $\Phi_{\rm{SB}}$ at the metal-InSb NW interfaces.
Here the device has a large contact spacing $L_{\rm{c}}\sim1.2\ \rm{{\mu}m}$ and
thus operates in the diffusive regime. %
Figure~2a shows the two-probe conductance $G$ of the device,
obtained by dividing the measured S-D current $I_{\rm{ds}}$ by $V_{\rm{ds}}$, as a function of the applied gate voltage $V_{\rm{gs}}$ at different temperatures ($T$).
The measurements were performed at a constant $V_{\rm{ds}} = +0.1$ V.
The flat-band voltage $V_{\rm{FB}}$ and the threshold voltage $V_{\rm{th}}$ of this device
at $T = 13$ K are 11 V and 16.4 V , respectively.
As shown in the figure, at high temperatures, the device shows a high off-state conductance and a low on/off conductance ratio.
At temperatures below 150 K,  the device shows an asymmetric ambipolar behavior.
This is a typical signature of a narrow band gap NWFET~\cite{Zhao2012}.
As has been discussed by Appenzeller \textit{et al.}~\cite{Appenzeller2004}, the presence of SBs affects largely the transport characteristics of a NWFET device in the gate voltage region close to the pinch-off threshold.
For $V_{\rm{gs}} > V_{\rm{th}}$, the injecting current consists of a thermionic emission component $I_{\rm{thermo}}$ and
a direct tunneling component $I_{\rm{tunnel}}$ (see Figure 2c, inset I).
For $V_{\rm{FB}} < V_{\rm{gs}} < V_{\rm{th}}$, it is the thermally-assisted tunneling through the barrier that dominantly modifies the current injection.
The crossover from a tunneling-modified to a thermionic emission-dominated process occurs at the flat-band condition, $V_{\rm{gs}} = V_{\rm{FB}}$ (Figure~2c, inset II). As continuously decreasing $V_{\rm{gs}}$, the thermionic emission over the barrier becomes the main, if not only, access for current injection.
To extract the true $\Phi_{\rm{SB}}$, the transfer curves in Figure~2a are converted to the Arrhenius plots  at different gate voltages shown in Figure~2b.
The data at high temperatures can then be fitted to a classical thermionic emission equation, to extract the effective barrier height $\Phi_{\rm{B}}$, as
\begin{equation}
I_{\rm{ds}}=A^{*}T^{2}\exp(-e\Phi_{\rm{B}}/k_{\rm{B}}T)[1-\exp(-eV_{\rm{ds}})/k_{\rm{B}}T],
\end{equation}
where $A^{*}$ is the Richardson constant, $k_{\rm{B}}$ is the Boltzmann constant, and $e$ is the elementary charge. The results are shown in Figure~2c.
It is seen that the extracted $\Phi_{\rm{B}}$ is modulated by $V_{\rm{gs}}$. Based on the
above band alignment analysis, $\Phi_{\rm{B}}$ is linearly related to $\Phi_{\rm{SB}}$ as
$\Phi_{\rm{B}}=\Phi_{\rm{SB}}+\beta(V_{\rm{gs}}-V_{\rm{FB}})$ for $V_{\rm{gs}}$ $\leq$ $V_{\rm{FB}}$, where $\beta$ is a voltage scaling factor.
We extract a Schottky barrier height of $\Phi_{\rm{SB}}\approx20\ \rm{meV}$ from the flat-band condition. Note that eq (1) works only for a single barrier case.
Considering the small $\Phi_{SB}$ extracted, the applied +0.1 V S/D bias
is large enough to eliminate the SB at the drain contact,
while avoids possible band-to-band tunneling current injection at higher biases.
As $V_{gs}$ is continuously decreased, $\Phi_{\rm{B}}$ shows a decrease as seen in the left-side region of Figure~2c.
This can be understood as the emergence of hole current, as denoted by a red arrow in the inset III of Figure~2c, when the Fermi level is positioned close to the valence band.
Accordingly, $\Phi_{\rm{B}}$ reaches a maximum when the Fermi level at the metal contact is in lineup with the middle of the  band gap. This simple scenario yields an approximate upper-bound value of the NW band gap $E_{\rm{g}}\sim\ 200\ \rm{meV}$.
We should note that the extracted values of $\Phi_{\rm{B}}$ become negative at the on state of the device as shown in Figure~2c, which indicates that the method for extraction of $\Phi_{\rm{B}}$ is no longer applicable for the device at on state. The negative values of $\Phi_{\rm{B}}$ arise from the fact that at on state, the contact resistance of the NWFET is relatively small and the measured temperature dependence of the current in the two-probe configuration is dominantly induced by the change of the nanowire resistance with varying temperature. To see this, we show in Figure~2d the measured on-state resistance $R_{\rm{on}}$ of the device with decreasing temperature. The data points in green circles are extracted by taking into account a $T$-dependent shift of the threshold voltage $V_{\rm{th}}$ as shown in the inset of Figure~2d and setting $V_{\rm{gs}}$ at values of 12 $\rm{V}$ higher than $V_{\rm{th}}$.
As a comparison, the $R_{\rm{on}}$ values measured at a fixed high value of $V_{\rm{gs}}$=29 $\rm{V}$) are also plotted in Figure~2d as blue circles. In both cases, $R_{\rm{on}}$  show an overall decrease with decreasing temperature, which is seldom investigated in SB height extraction experiments~\cite{Appenzeller2004,Razavieh2014,Das2013}.
The decrease of $R_{\rm{on}}$, or equivalently the increase of $I_{\rm{on}}$, with lowering $T$ has been reported before in InAs NWFETs in both the ballistic and the diffusive transport regime~\cite{Chuang2013,Wang2015} and can be explained by suppression of phonon scattering at low temperatures.

\begin{figure*}

  \centering
  \includegraphics[width=6in]{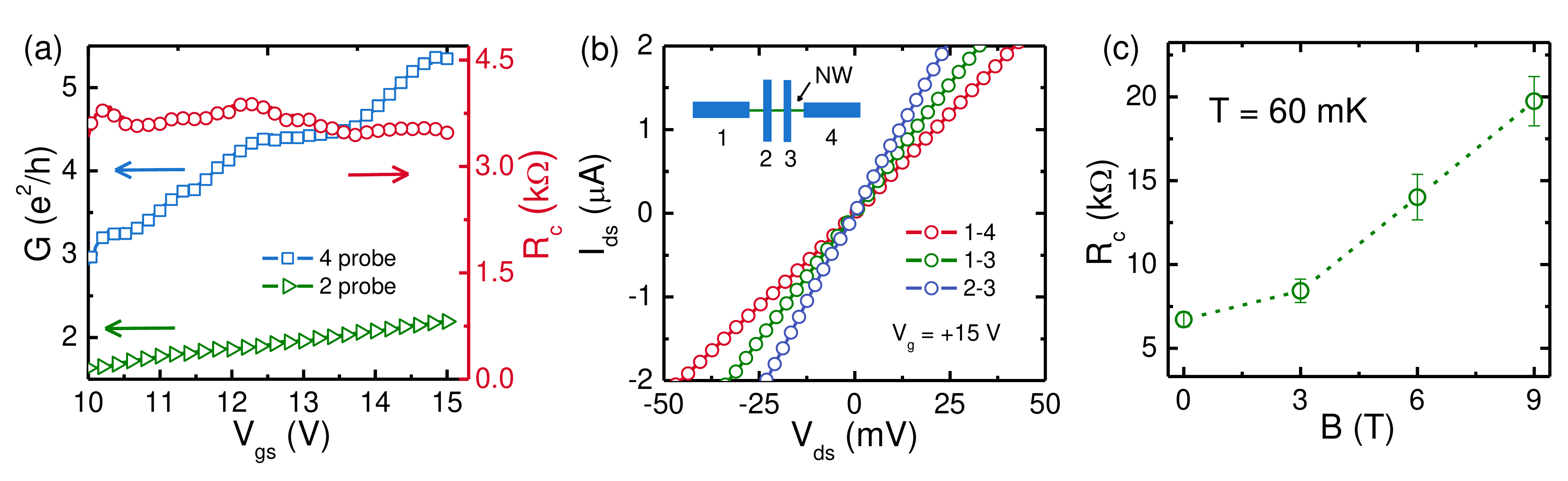}
  \caption
  {Contact resistance $R_{\rm{c}}$ at low temperatures and its dependence on magnetic field. (a) Representative two-probe conductance (green) and four-probe conductance (blue) measured for a four-probe device (Device 2) at on state as a function of gate voltage $V_{\rm{gs}}$ at a temperature of $T=60$ mK. The two-probe contact resistance is measured between probes 2 and 3 [see the schematic shown in (b)]. Red circles show the contact resistance $R_{\rm{c}}$, extracted from the measured two-probe and four-probe conductance of the device.  (b) Source-drain current $I_{\rm{ds}}$ measured between different probes of the device as a function of $V_{\rm{ds}}$ at $V_{\rm{gs}}= 15\ \rm{V}$ (on state) and $T=60$ mK. The inset shows a sketch of the device with the probes labeled. The contact spacings between probes 1 and 4, probes 1 and 3, and probes 2 and 3 are 1240, 740, and 260 nm, respectively. (c) Contact resistance $R_{\rm{c}}$ extracted for a four-probe device (Device 3) at on state as a function of magnetic field $B$ at $T=60$ mK. }
  \label{Figure 3}
\end{figure*}

Now we turn to consider the contact resistance $R_{\rm{c}}$ at the Ti/Au-InSb NW interfaces.
We can reliably extract the contact resistance $R_{\rm{c}}$ from measurements of a four-probe device as shown in Figure~1b at low temperatures~\cite{Mohney2005}.
Figure~3a shows the results of the measurements for a typical four-probe device (Device 2) at on state at $T=60\ \rm{mK}$. In these measurements,  a two-probe $G-V_{\rm{gs}}$ trace is first recorded using the two inner electrodes 2 and 3, while keeping the two outer electrodes 1 and 4 floated (\textit{cf.} Figure~1b or the inset of Figure~3b). Then a four-probe $G-V_{\rm{gs}}$ trace is taken by applying a constant current using electrodes 1 and 4. The contact resistance can be determined from the measured two-probe and four-probe conductance according to $R_{\rm{c}}=(G_{\rm{2p}}^{-1}-G_{\rm{4p}}^{-1})/2$, where $G_{\rm{2p}}$ and $G_{\rm{4p}}$ represent the two-probe and the four-probe conductance, respectively.
For Device 2 at on state, $G_{\rm{4p}}$ is around 2 times of $G_{\rm{2p}}$, giving an average contact resistance of $R_{\rm{c}}=3.63\pm0.18\ \rm{k\Omega}$ and a normalized value of $1.20\pm0.06\ \rm{k\Omega\cdot\mu m}$ (normalized to the NW circumference).
Altogether, three four-probe devices are measured and it is found that $R_{\rm{c}}$ shows device-to-device variations in a range of $3.63-6.81\ \rm{k\Omega}$. It is worthwhile to note that at on state, the measured $I_{\rm{ds}}-V_{\rm{ds}}$ characteristics of our devices all show good ohmic behaviors. An example is shown in Figure~3b, where the $I_{\rm{ds}}-V_{\rm{ds}}$ characteristics of Device 2 at on state measured using different combinations of two probes at $T=60\ \rm{mK}$ are presented. Here, it is seen that all the measured curves are linear and ohmic like.


To analyze the physical origin of the contact resistance $R_{\rm{c}}$ of InSb NWFETs at low temperatures, we carry out the magnetotransport measurements and examine the effect of magnetic field $B$ on the extracted $R_{\rm{c}}$ of our devices. Figure~3c shows the magnetic field dependence of $R_{\rm{c}}$ measured for a four-probe device (Device 3) at on state. Here, the magnetic field is applied perpendicular to the substrate and each $R_{\rm{c}}$ value is obtained by averaging over a $V_{\rm{gs}}$ range of $5\ \rm{V}$ at on state with standard errors indicated by vertical bars.
Interestingly, $R_{\rm{c}}$ develops a pronounced upturn at high fields.
The transition to the upturn in the  $R_{\rm{c}}-B$ curve can be roughly estimated to occur around $3\ \rm{T}$.
To understand this transition, we consider qualitatively the response of the band structure of the NW to a change in the applied magnetic field.
Generally, the contact resistance $R_{\rm{c}}$ contains two terms, $R_{\rm{b}}$ and $R_{\rm{q}}$. The first term, $R_{\rm{b}}$, comes from the potential barrier at the metal-NW interface and is sample-dependent due to specific interface properties. The second term, $R_{\rm{q}}$, is due to the mode mismatch between three-dimensional electron states in the metal and quasi-one-dimensional subband states in the semiconductor NW. At on state, it is reasonably assumed that the Schottky barrier at the metal-semiconductor NW interface is very thin and is transparent for electron conduction. Thus,
$R_{\rm{b}}$ can be neglected and we have approximately $R_{\rm{c}}\approx R_{\rm{q}}$ for the NWFET at on state at low temperatures. Under this condition, $R_{\rm{c}}$ can be analyzed based on the well-known Landauer formula,
$G=\frac{2e^{2}}{h}\sum\limits_{i=1}^{N}T_{\rm{i}}$, where $N$ is the number of conduction modes, $T_{\rm{i}}$ is the transmission coefficient of each mode, and $h$ is the Planck constant~\cite{Datta1995}.
We have estimated, from the low-temperature transfer characteristics of the device, the  electron mean free path in the NW is about $30-50\ \rm{nm}$, which is much smaller than the spacing of $\sim250\ \rm{nm}$ between the probes. Thus, our device is in the quasi-ballistic or the diffusive transport regime. Yet, the quantized conductance have been reported in InSb \cite{Weperen2012} and InAs~\cite{Chuang2013} NWs with contact spacings several times larger than the mean free path, no such signatures have been observed in the devices reported here. Nevertheless, the transport in our devices is still sensitively dependent on the magnetoelectric subbands in the NW at an applied magnetic field.
As the field is increased, various features are expected to appear, such as a larger subband separation, flattening of the energy band inside the NW, and successive depopulation of the magnetoelectric subbands~\cite{Knobbe2005,Nanot2009,Vigneau2014}. In the context of transport, the depopulation of the magnetoelectric subbands amounts to a fall in the number of conducting channels below the Fermi level at high magnetic fields, or equivalently, the occurrence of upturn in the contact resistance as observed in our experiment. The relevant length scale for the onset of the magnetic field dependence of $R_{\rm{c}}$ is the cyclotron orbit radius $l_{\rm{c}}$, which can be estimated from the form of $l_{\rm{c}}=k_{\rm{F}}l_{\rm{B}}^{2}$, where $k_{\rm{F}}$ is the Fermi wavevector and $l_{\rm{B}}=(\hbar/eB)^{1/2}$ is the magnetic length with $\hbar$ being the reduced Planck constant. When the cyclotron orbit radius becomes $l_{\rm{c}}\approx d/2$ or smaller at high magnetic field, the Landau levels are formed in the inside of the NW and the magnetoelectric subbands bend up at the NW boundary, leading to the formation of edge state channels~\cite{Royo2013}. As a consequence, both the channel mobility in the NW can be increased due to the suppression of back scattering and the contact resistance $R_{\rm{q}}$ can be increased due to the depopulation of magnetoelectric subbands. The above analysis gives the estimation for the onset field $B_{\rm{c}}$, at which the contact resistance $R_{\rm{q}}$ starts to increase with increasing magentic field, as
\begin{equation}
\centering
B_{\rm{c}}=\frac{2\hbar k_{\rm{F}}}{ed}.
\end{equation}
For Device 3, using the $k_{\rm{F}}$ determined from the transfer characteristics, an onset field of $B_{\rm{c}}\sim2.8\ \rm{T}$ can be deduced, which is close to the measured value of $3\ \rm{T}$.

\begin{figure}

  \centering
  \includegraphics[width=2.8in]{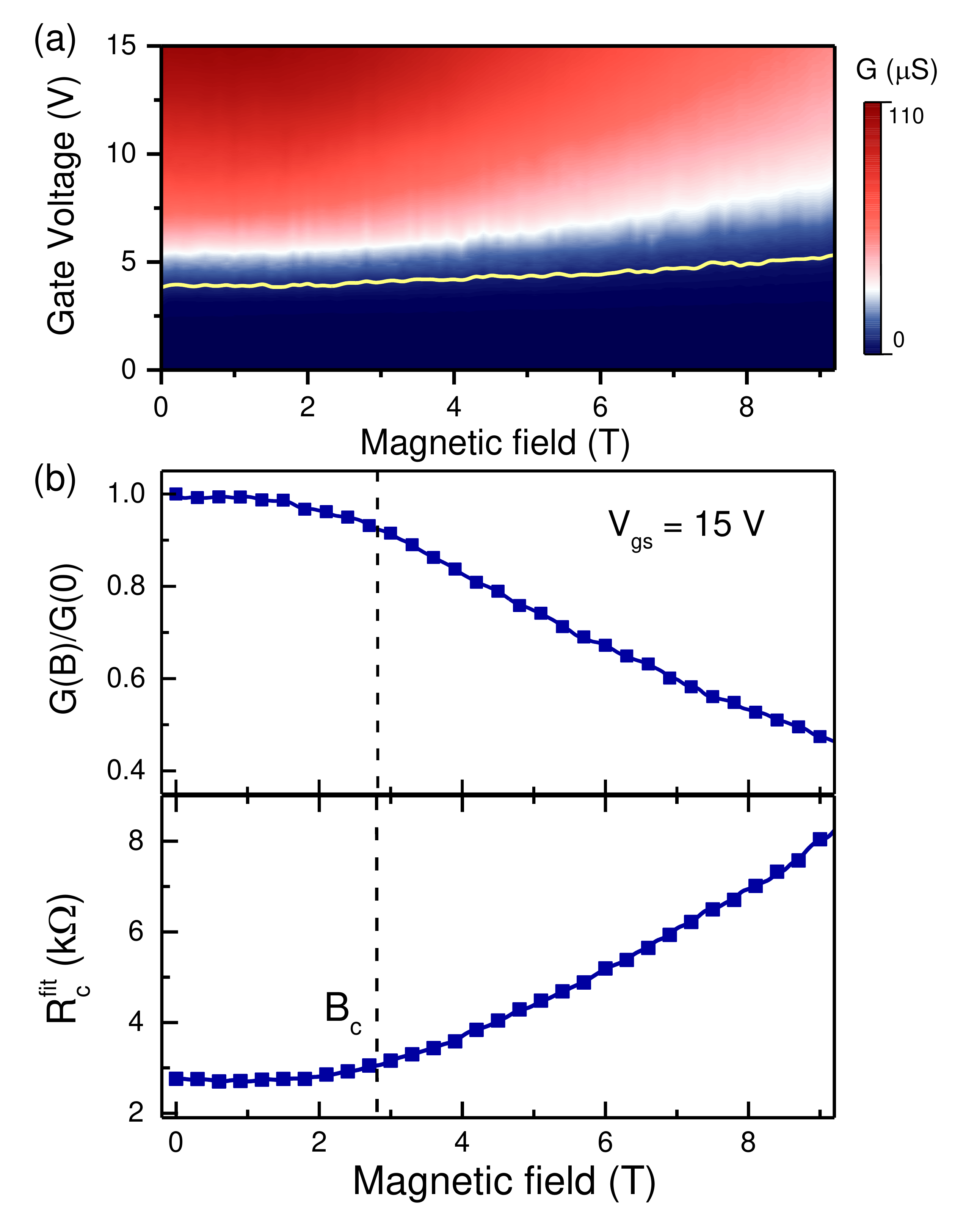}
  \caption
  {(a) Conductance $G$ measured as a function of gate voltage $V_{\rm{gs}}$ and magnetic field $B$ for a two-probe device (Device 4) at source-drain voltage $V_{\rm{ds}}=0.1\ \rm{V}$ and temperature $T=60\ \rm{mK}$. The dark blue (dark red) color corresponds to a low (high) value of the conductance. The yellow line denotes the pinch-off threshold voltages of the device at different values of $B$. (b) Normalized conductance $G(B)/G(0)$ (upper panel) measured and  contact resistance $R_{\rm{c}}^{\rm{fit}}$ (lower panel) extracted by the fitting procedure as described in the main text for the device at gate voltage $V_{\rm{gs}}=15$ V (on state) and temperature $T=60$ mK as a function of $B$. The two graphs share a common transition field of $B_{\rm{c}}\sim 2.8\ \rm{T}$.}
  \label{Figure 4}
\end{figure}

To further support the scenario of the magnetic depopulation of subbands in InSb NWs, we carried out magnetoconductance measurements on another two-probe device (Device 4). Figure~4a shows the measured conductance $G$ of the device as a function of $V_{\rm{gs}}$ and $B$. The yellow line in the figure denotes the threshold gate voltage at which the NW channel is open for conduction. It is evident that the threshold gate voltage is increased with increasing $B$. The upper panel of Figure~4b shows the measured conductance $G$ (normalized to the conductance at $B=0$) of the device at on state as a function of $B$. Here, an overall descending trend of the conductance with increasing magnetic field is seen. This seems at first sight counter-intuitive given the fact that the introduction of a magnetic field suppresses back scattering in the NW and thus leads to an increase in the channel mobility. Indeed, Dhara \textit{et al.} observed a mobility increase in the magnetotransport measurements of InAs NWs~\cite{Dhara2009}. In our case, however, because of the presence of the contact resistance, the field-effect mobility $\mu_{\rm{FE}}$ can not be directly extracted from two
terminal measurements with the expression $\mu_{\rm{FE}}=g_{\rm{m}}L_{\rm{c}}/C_{\rm{g}}$, where $g_{\rm{m}}=dG/dV_{\rm{gs}}$ is the maximum transconductance, $L_{\rm{c}}$ is the contact spacing and $C_{\rm{g}}$ is the gate-to-nanowire capacitance per unit length \cite{Deyeh2007}. Instead, we need to adopt the method reported by G\"{u}l \textit{et al.}~\cite{Gul2015} by incorporating $R_{\rm{c}}$ into the expression of the conductance $G$ as follows,
\begin{equation}
G(V_{\rm{gs}})=[R_{\rm{c}}+\frac{L_{\rm{c}}}{\mu_{\rm{FE}}C_{\rm{g}}(V_{\rm{gs}}-V_{\rm{th}})}]^{-1}.
\end{equation}
Here $R_{\rm{c}}$, $\mu_{\rm{FE}}$, and $V_{\rm{th}}$ are free fitting parameters. In analysis of our measured results, we assume that $R_{\rm{c}}$ is magnetic field dependent and $\mu_{\rm{FE}}$ is independent of the magnetic field, i.e., neglecting the effect of the increases in $\mu_{\rm{FE}}$ with increasing magnetic field.  The lower panel in Figure~4b shows the result of $R_{\rm{c}}^{\rm{fit}}$ obtained by fitting the measured data shown in Figure~4a to the above expression as a function of $B$. Strikingly, we noticed a shared feature for the two panels---both the normalized conductance $G$ and the extracted contact resistance $R_{\rm{c}}^{\rm{fit}}$ first undergo an initial stage with small variations and then start to decrease  or increase quickly after $B \sim 3\ T$, a magnetic field value which is very close to the onset field $B_{\rm{c}}\sim 2.8\ T$ deduced for this device (dashed line). This consistency is not a coincidence but reveals unambiguously the critical role of the contact resistance arising from mode mismatching. The significant rise in $R_{\rm{c}}$ after the applied magnetic field is stronger than $B_{\rm{c}}$ provides a strong hint on the subband structure modulation by the magnetic field and thus supports our proposed dominant physical origin of the contact resistance of InSb NWFETs at on state at low temperatures.


\section{Conclusions}

In conclusion, we have carried out an experimental study of Ti/Au-contacted InSb NWFETs with two-probe and four-probe designs and have extracted the Schottky barrier height and contact resistance at the Ti/Au-InSb NW interfaces. We have shown that a Schottky barrier height of $\Phi_{SB}\sim 20\ \rm{meV}$ is present at the Ti/Au-InSb NW interfaces and the effective barrier height is gate tunable. We have also demonstrated that the contact resistance  $R_{\rm{c}}$ of the devices at on-state arises dominantly from mode mismatching between the electron states in the contact metal and the semiconductor NWs and is magnetic field dependent. The significant increase of $R_{\rm{c}}$ after an onset magnetic field is interpreted as due to the magnetic depopulation of the subbands in the NWs and the onset field strength can be estimated simply by setting the electron cyclotron radius to a half of the NW diameter. Our finding of a variable effective Schottky barrier height points to readily feasible device designs, where the injection barrier can be tuned by local gates, aiming at both applications and fundamental research.


\ack{The authors gratefully acknowledge Mengqi Fu for fruitful discussions. This work was supported by the National Basic Research Program of the Ministry of Science and Technology of China (Grant Nos. 2012CB932703 and 2012CB932700), the National Natural Science Foundation of China (Grant Nos. 91221202, 91421303, 61321001, and 11374019), and the Swedish Research Council (VR). N~K thanks also the Ph.D. Program Foundation of the Ministry of Education of China for financial support (Grant No. 20120001120126). }

\section*{References}
\bibliography{references}

\end{document}